\documentclass[lineno]{jfm}
\usepackage{color,graphicx,array,tabularx,bm,url}
\usepackage{epstopdf,epsfig}
\usepackage{bm}
\usepackage{amsmath}
\usepackage{newtxtext}
\usepackage{newtxmath}
\usepackage{natbib}
\usepackage{hyperref}

\title{Instability of a dusty Kolmogorov flow}

\author{
Alessandro Sozza\aff{1}\aff{2}\aff{3}\corresp{\email{asozza.ph@gmail.com}}, 
Massimo Cencini\aff{2},
Stefano Musacchio\aff{1} 
and Guido Boffetta\aff{1}
}

\affiliation{
\aff{1}Department of Physics and INFN, University of Torino, via P. Giuria 10125, Torino, Italy.
\aff{2}Istituto dei Sistemi Complessi, ISC-CNR, via dei Taurini 19, 00185, Roma, Italy and INFN sez. Roma2 "Tor Vergata".
\aff{3}Laboratoire de Physique, UMR 5672, \'Ecole Normale Sup\'erieure de Lyon, 46 All\'ee d'Italie, 69007 Lyon, France.}

\begin{document}

\maketitle

%%%%%%%%%%%%%%%%%%%%%%%%%%%%%%%%%%%%%%%%%%%%
\begin{abstract}
Suspended particles can significantly alter the fluid properties and, 
in particular, can modify the transition from laminar to turbulent flow. 
We investigate the effect of heavy particle suspensions on the linear 
stability of the Kolmogorov flow by means of a multiple scale expansion 
of the Eulerian model originally proposed by \citet{saffman1962}. 
We find that, while at small Stokes numbers particles always destabilize 
the flow (as already predicted by Saffman in the limit of very thin particles), 
at sufficiently large Stokes numbers the effect is non-monotonic in the 
particle mass fraction and particles can both stabilize and destabilize the flow.
Numerical analysis is used to validate the analytical predictions.
We find that in a region of the parameter space the multiple-scale 
expansion overestimates the stability of the flow and that this is a consequence 
of the breakdown of the scale separation assumptions.
\end{abstract}
%%%%%%%%%%%%%%%%%%%%%%%%%%%%%%%%%%%%%%%%%%%%

%\begin{keywords}
%\end{keywords}

%{\bm MSC Codes } 

%%%%%%%%%%%%%%%%%%%%%%%%%%%%%%%%%%%%%%%%%%%%
%%%%%%%%%%%%%%%%%%%%%%%%%%%%%%%%%%%%%%%%%%%%
\section{Introduction}
\label{sec:intro}

Particles transported in flow are ubiquitous in many natural environment, 
from protoplanetary disks \citep{armitage2011dynamics}, to aerosol 
in the atmosphere \citep{shaw2003particle}, from volcanic eruptions 
\citep{bercovici2010} to sediment transport \citep{burns2015}.

Dispersed particles are not only transported by the flow, but they exert 
forces on the fluid that, depending on the mass loading, can modify the flow itself.
As discovered long ago \citep{sproull1961viscosity}, at high Reynolds 
number heavy particles can alter turbulence by attenuating or enhancing it 
depending on their size and mass fraction and on the scale considered 
\citep{balachandar2010,gualtieri2017turbulence,bec2017}. 
In channel flow, they can change the turbulent drag 
\citep{li2019drag,ardekani2017}. At low Reynolds numbers, 
the presence of particles affects the stability of laminar flow and 
the transition to turbulence. Indeed, as first realized by 
\citet{saffman1962}, tiny particles, characterized by small Stokes 
number, typically anticipate the onset of the instability while 
coarser ones retard it. This intuition was later confirmed by other 
studies in the context of pipe 
\citep{michael1964stability,rudyak1997hydrodynamic} and channel 
\citep{klinkenberg2011} flows. However, in wall bounded flows the 
analysis is complicated by the interaction of particles with the 
boundaries and by the fact that the transition is subcritical and thus 
finite amplitude perturbations are required to destabilize the flow.

In this work we study the effects of a particle suspension on the
stability of a periodic Kolmogorov flow. This sinusoidal flow was
proposed by Kolmogorov as a simple model to understand the transition
to turbulence and, after seven decades of studies, 
it is still a attracting a broad scientific interest (for a recent review see, 
e.g., \citet{fylladitakis2018kolmogorov}). From a theoretical point of view, it
has the advantage, with respect to other parallel shear flows, 
to be analytically tractable for studying its linear stability and 
weakly non-linear dynamics \citep{sivashinsky1985weak}. 
In numerical simulations, it is widely used as a 
prototype of shear flow with periodic boundary conditions which 
can be easily implemented in pseudo-spectral codes. 
Moreover, the Kolmogorov flow can be considered as a simplified 
channel flow without boundaries, since it displays a mean velocity 
profile which remains monochromatic even in the turbulent regime 
\citep{musacchio2014}. For these reasons, analytical and numerical 
studies have extended the Kolmogorov flow to the $\beta$-plane \citep{legras1999}, 
to stratified \citep{balmforth2002} and viscoelastic flows \citep{boffetta2005}.
We remind that, beside the numerical and analytical  studies, 
the Kolmogorov flow is also realizable in experiments \citep{suri2014velocity}. 
Recently, the Kolmogorov flow has been also used to study numerically the 
clustering of inertial particles \citep{de2016clustering,pandey2019clustering} 
as well as the effect of an heavy particle suspension on turbulent drag 
\citep{sozza2020drag}. The latter numerical study has been performed 
by using an Eulerian approach originally developed by \citet{saffman1962}, 
valid in the limit of small volume fraction for mono-disperse heavy particle suspensions.

In the present work we consider the laminar stationary solution of the 
Saffman model forced by a Kolmogorov flow. We show that it is possible 
to study the stability problem perturbatively, by exploiting a multiple-scale 
expansion \citep{bensoussan2011asymptotic}. The analytical result, which 
extends the Newtonian one \citep{sivashinsky1985negative}, predicts a rich 
phenomenology with both enhanced and reduced stability as a function of 
the control parameters, namely the particle Stokes number and mass fraction. 
In particular, we confirm the known phenomenology that tiny (coarse) particles 
tend to destabilize (stabilize) the flow with respect to the Newtonian case. 
Moreover, we show that for coarse enough particles the effect is 
non-monotonic in the mass fraction: at small mass fractions the flow is 
stabilized while it is destabilized at large enough mass fractions. 
A similar phenomenology was observed for neutrally buoyant particles 
in pipe flows \citep{matas2003,agrawal2019}. We compare the analytically 
predicted critical Reynolds number with the results of an extended 
numerical investigation and we explain the observed discrepancies 
for some values of the parameters with the 
breakdown of the scale separation assumption.

The remaining of this paper is organized as follows. 
In Section~\ref{sec:2} we introduce the Saffman model. 
In Section~\ref{sec:3} we perform the linearization 
around the Kolmogorov base flow. 
Section~\ref{sec:4} is devoted to the multiple-scale 
approach for the linear stability problem. 
In Section~\ref{sec:5} we discuss the dependence of the 
critical Reynolds number of the control parameters and compare 
the analytical predictions with the numerical results. 
Finally, Section~\ref{sec:conclusions} is devoted to conclusions.
%%%%%%%%%%%%%%%%%%%%%%%%%%%%%%%%%%%%%%%%%%%%

%%%%%%%%%%%%%%%%%%%%%%%%%%%%%%%%%%%%%%%%%%%%
\section{Saffman model for a dusty Kolmogorov flow}
\label{sec:2}

We consider an Eulerian model for a dilute suspension of heavy particles
with two-way coupling introduced by \citet{saffman1962} long ago.
The model considers a dilute mono-disperse suspension of small, 
heavy, spherical particles with density $\rho_p$ and radius $a$ 
transported by a Newtonian fluid with density $\rho_f$ and viscosity $\mu$. 
Particle size is assumed to be much smaller than any scale in the 
flow such that the particle Reynolds number is negligible. 
The particle volume fraction $\phi_v = N_p v_p / V$, defined in terms 
of the volume of each particle $v_p=4\pi a^3/3$ and the number 
of particles $N_p$ contained in the total volume $V$, is assumed 
to be negligible while the mass fraction $\phi=\phi_v \rho_p/\rho_f$ 
can be of order one since it is assumed $\rho_p \gg \rho_f$ 
(as in a dilute suspensions of water droplets in air).

Within the model, the fluid density field remains constant because of 
the assumption of vanishing $\phi_v$ and it is transported by the 
incompressible velocity field of the fluid phase ${\bm u}(\bm x,t)$. 
The solid phase is described by a number density field
$\theta(\bm x,t) = n(\bm x,t) /(N_p/V)$,  
where $n(\bm x,t)$ is the local number of particles per unit volume.  
The normalization gives $\langle \theta \rangle = 1$, 
where the brackets $\langle [\cdot] \rangle$ denote the average over 
the volume $V$.  
The number density field $\theta$ is transported by a compressible 
particle velocity field ${\bm v}(\bm x,t)$. 

For small volume fractions ($\phi_v < 10^{-3}$) the dynamics of the 
particle-laden flow can be described by a two-way coupling, which 
takes into account the interactions between individual particles and 
the surrounding flow, but neglects the interactions between particles 
(collisions and friction) and the particle-fluid-particle interactions 
(fluid streamlines compressed between particles) 
\citep{elghobashi1994predicting}. In the two-way coupling regime, 
the exchange of momentum between the two phases can no longer 
be neglected \citep{balachandar2010}. For small heavy particles, 
such an exchange is mainly mediated by the viscous drag force which 
is proportional to the difference between particle and fluid velocities. 

The accurate modeling of the coupling between the 
particles and the flow is a challenging task. In Lagrangian-Eulerian 
approaches based on the point-particle method, it requires to take 
into account the local perturbation to the fluid due to the presence 
of the particle \citep{horwitz2016accurate}. 
The Eulerian model proposed by Saffman is based on a simplified 
assumption, namely that the coupling is obtained by imposing the 
local conservation of the total momentum of the fluid and particle 
phases. This leads to the following equations \citep{saffman1962}:
\begin{eqnarray}
\partial_t {\bm u} + {\bm u}\cdot{\bm \nabla}{\bm u} &=& 
- {\bm \nabla}p + \nu \nabla^2 {\bm u} + {\bm f} + \frac{\phi}{\tau}\theta ({\bm v} - {\bm u})
\label{eq:u}\\
\partial_t {\bm v} + {\bm v}\cdot{\bm \nabla}{\bm v} &=& 
- \frac{1}{\tau} ({\bm v} - {\bm u})
\label{eq:v}\\
\partial_t \theta + {\bm \nabla}\cdot \left({\bm v}\theta\right) &=& 0
\label{eq:theta}
\end{eqnarray}
where $\tau = (2/9) a^2 \rho_p/(\rho_f \nu)$ is the relaxation time of the particles, 
$\nu=\mu/\rho_f$ is the kinematic viscosity and ${\bm f}$ is an external forcing.

In the limit of very tiny particles, i.e. small $\tau$, the Saffman model 
reduces to the Navier-Stokes equation for an incompressible flow 
with an increased density, and thus a smaller viscosity~\citep{saffman1962}. 
Indeed when $\tau \to 0$ from (\ref{eq:v}) one has ${\bm v}={\bm u}$.
For small $\tau$ one can expand ${\bm v}={\bm u}+\tau \tilde{\bm v}+O(\tau^2)$
and (\ref{eq:v}) gives, at leading order,
$\tilde{\bm v}=-(\partial_t {\bm u}+{\bm u}\cdot{\bm \nabla}{\bm u})+O(\tau)$.
Substituting now
${\bm v}={\bm u} -\tau (\partial_t {\bm u}+{\bm u}\cdot{\bm \nabla}{\bm u})$
in Eq.~((\ref{eq:theta}) one 
obtains that the particle density field remains constant at leading 
order. Finally, using $\theta=1+O(\tau)$ and 
$({\bm v}-{\bm u})/\tau = -(\partial_t {\bm u}+{\bm u}\cdot{\bm \nabla}{\bm u}) + O(\tau)$ 
in (\ref{eq:u}) gives:
\begin{equation}
\partial_t {\bm u} + {\bm u}\cdot{\bm \nabla}{\bm u} = 
- {\bm \nabla}p + \frac{\nu}{1+\phi} \nabla^2 {\bm u} + \frac{{\bm f}}{1+\phi}\,,
\label{eq:smalltau}
\end{equation}
i.e. the Navier-Stokes equation for an incompressible velocity field 
with forcing and viscosity rescaled by the factor $(1+\phi)$.

Remarkably, we show that the same result is also recovered in the 
limit of large $\phi$. Indeed, from Eq.~(\ref{eq:u}) one can write
\begin{equation}
{\bm u} = {\bm v} + \frac{\tau}{\phi} \frac{1}{\theta}
\left(-\partial_t {\bm u} + {\bm u} \cdot {\bm \nabla u} - 
{\bm \nabla} p + \nu \nabla^2 {\bm u} + {\bm f} \right)\,,
\label{eq:largephi1}
\end{equation}
showing that the difference between ${\bm u}$ and ${\bm v}$ is 
of order $1/\phi \ll 1$. Substituting ${\bm v}={\bm u} + O(1/\phi)$ 
in Eq.~(\ref{eq:theta}) implies $\theta=1+O(1/\phi)$ which, 
together with Eq.~(\ref{eq:largephi1}) in Eq.~(\ref{eq:v}), yields 
Eq.~(\ref{eq:smalltau}) multiplied by $(1+\phi)$, 
i.e. again a Navier-Stokes equation with rescaled forcing and 
viscosity.
We remark that the limit of large $\phi$ is physically 
questionable since it could violate the assumption of negligible 
volume fraction. Nonetheless, it is mathematically well defined 
and we will use it the following to discuss our results.

We consider the case of a monochromatic periodic forcing 
${\bm f} = F \cos (K y) \widehat{\bm x}$ which produces the 
Kolmogorov laminar fixed point 
${\bm u}({\bm x})={\bm v}({\bm x})={\bm U}(y) \equiv U \cos(K y)$ 
with $U=F/(\nu K^2)$ and $\theta({\bm x})=1$. 
We remark that in general the Kolmogorov flow is a stationary 
solution also for $\theta({\bm x})=g(y)$ with $g$ arbitrary function. 
Nonetheless, the solution with uniform density $\theta$ is 
physically the more relevant as it survives to the presence 
of an arbitrarily small diffusivity. 

The non-dimensional parameters of the model are the Reynolds number 
$Re=U/(\nu K)$, defined in terms of the amplitude of the laminar flow $U$
and on the only characteristic length of the flow $K^{-1}$,
the Stokes number $St=\tau \nu K^2$, defined as the ratio between the 
particle relaxation time $\tau$ and the viscous time $\tau_\nu=1/(\nu K^2)$, 
and the mass fraction $\phi$. In the following, we will study the linear stability 
of the laminar fixed point as a function of $Re$, $St$ and $\phi$.

We conclude this Section with a comment about the limitations 
of the Saffman model.  Beside the assumption of small volume 
fraction, in the case of turbulent flows at high Reynolds numbers 
the validity of the model (\ref{eq:u}-\ref{eq:theta}) is in general 
limited to small Stokes numbers $St < 1$. This is due to the 
phenomenon of caustics \citep{wilkinson2005} which would imply 
a multi-valued particle velocity field breaking the validity of the 
continuum description. Nonetheless, for the specific case of the 
linear stability of a laminar parallel flow considered here, in the 
laminar fixed point the particle velocity field is equal to the fluid 
velocity field, and therefore the model is well defined for arbitrary 
value of $St$.

%%%%%%%%%%%%%%%%%%%%%%%%%%%%%%%%%%%%%%%%%%%%
\section{Linear stability analysis}
\label{sec:3}

We study the linear stability of an infinitesimal perturbation 
of the basic Kolmogorov flow. To this aim we expand 
Eqs.~ (\ref{eq:u}-\ref{eq:theta}) around the laminar fixed point 
${\bm u}({\bm x},t)={\bm U}(y)+{\bm u}'({\bm x},t)$, 
${\bm v}({\bm x},t)={\bm U}(y)+{\bm v}'({\bm x},t)$, 
$\theta({\bm x},t)=1+\theta'({\bm x},t)$, 
and obtain the linearized equations for the perturbations: 
\begin{eqnarray}
\partial_t {\bm u}' + {\bm U}\cdot{\bm \nabla}{\bm u}' +
{\bm u}'\cdot{\bm \nabla}{\bm U}&=&
- {\bm \nabla}p' + \nu \nabla^2 {\bm u}' + 
\frac{\phi}{\tau}({\bm v}' - {\bm u}')
\label{eq:u-lin}\\
\partial_t {\bm v}' + {\bm U}\cdot{\bm \nabla}{\bm v}' +
{\bm v}'\cdot{\bm \nabla}{\bm U}& =& 
- \frac{1}{\tau}({\bm v}' - {\bm u}')
\label{eq:v-lin}\\
\partial_t \theta' + {\bm U}\cdot{\bm \nabla}\theta'+
{\bm \nabla}\cdot {\bm v}'&=&0\,.
\label{eq:theta-lin}
\end{eqnarray}
We observe that, at this order, the density field becomes a passive 
scalar since it does not enter Eqs.~(\ref{eq:u-lin}-\ref{eq:v-lin}). 
Therefore, the evolution of $\theta'$ can be neglected.

A remarkable simplification of the linear stability analysis can be 
achieved by invoking the Squire's theorem for parallel flows 
\citep{squire1933}, which states that it suffices to consider 
two-dimensional perturbations, since three-dimensional perturbations
are more stable. From the original formulation, the theorem has been 
extended to various systems, including 
MHD equations \citep{hughes2001}, 
stratified flows \citep{balmforth2002}
and viscoelastic flows \citep{bistagnino2007}.
In the Appendix \ref{app:a}, we report the derivation of the Squire's 
theorem for the dusty fluid model 
(\ref{eq:u-lin}-\ref{eq:v-lin}).

In the following we will therefore consider the 
two-dimensional version of the linearized equation. It is convenient 
to rewrite the fluid velocity fluctuation in terms of a 
stream function ${\bm u}'=(\partial_y \Psi,-\partial_x \Psi)$ and the 
compressible particle velocity in terms of a particle stream function 
$\Psi_p$ and potential $\Phi_p$ as 
${\bm v}'=(\partial_y \Psi_p+\partial_x \Phi_p, -\partial_x \Psi_p+\partial_y \Phi_p)$. 
In terms of these fields the linear equations (\ref{eq:u-lin}-\ref{eq:v-lin}) read
\begin{eqnarray}
\strut\hspace{-0.9truecm}&&\partial_t \nabla^2 \Psi +U \cos(K y) (K^2+\nabla^2)\partial_x \Psi
-\nu \nabla^4 \Psi+ \frac{\phi}{\tau} \nabla^2 (\Psi \!-\!\Psi_p) \!=\!0
\label{eq:psi-lin}\\
\strut\hspace{-0.9truecm}&&\partial_t \nabla^2 \Psi_p \!+\!U \cos(K y) \left[\!(K^2\!+\!\nabla^2)\partial_x \Psi_p
\!-\!K^2 \partial_y \Phi_p \!\right]\!-\!U k \sin(K y) \nabla^2 \Phi_p \!+\! 
\frac{\nabla^2 (\Psi_p \!-\! \Psi)}{\tau} \!=\!0
\label{eq:psip-lin}\\
\strut\hspace{-0.9truecm}&&\partial_t \nabla^2 \Phi_p +U \cos(K y) \partial_x \nabla^2\Phi_p-
2 U K \sin(K y) \left(\partial_x \partial_y \Phi_p\! -\! 
\partial_x^2 \Psi_p \right)\!+\! \frac{1}{\tau} \nabla^2 \Phi_p\! =\!0\,.
\label{eq:phip-lin}
\end{eqnarray}

For a Newtonian fluid ($\phi=0$), the laminar solution is known 
to be linearly stable to perturbations at wavenumbers larger than $K$,
and to become unstable to large-scale transverse perturbations 
(i.e. in the direction $x$ transverse to the direction of modulation $z$) 
above the critical value $Re_c=\sqrt{2}$ \citep{meshalkin1961,sivashinsky1985negative}. 
As discussed in Sec.~\ref{sec:2}, 
in the limit of small inertia ($\tau \ll 1$) or large mass fraction ($\phi \gg 1$) 
the Saffman model recovers the Navier-Stokes equation with a rescaled 
viscosity $\nu/(1+\phi)$. Therefore, in these limits we expect the critical 
Reynolds number to become $Re_c = \sqrt{2}/(1+\phi)$, i.e. the presence 
of tiny particles, or a large mass fraction of particles,
makes the flow more unstable.

%%%%%%%%%%%%%%%%%%%%%%%%%%%%%%%%%%%%%%%%%%%%%%%%%%%%%%%
\section{Multiple scale analysis\label{sec:4}}

The general dependence of the critical Reynolds number on the
parameters $\tau$ and $\phi$ can be obtained by a standard
multiple-scale analysis~\citep{bensoussan2011asymptotic} of the
linearized equations (\ref{eq:psi-lin}-\ref{eq:phip-lin}).
The main idea of the multiple-scale method is to search for a 
perturbation which varies on spatial scales much larger than those 
of the base flow. For this purpose, beside the small-scale variables 
$x$,$y$ and $t$, the multiple scale method introduces the large-scale 
spatial variables $X=\varepsilon x$, $Y=\varepsilon y$ and a corresponding 
slow time $T=\varepsilon^2 t$, where the small parameter $\varepsilon$ 
is the ratio between the characteristic scales of the basic flow and the 
perturbation. The relative powers of $\varepsilon$ in the space and 
time variables reflect the diffusive dynamics expected at large scales. 
The two sets of variables are then assumed to be independent, so that by 
averaging over the small scales it is possible to obtain an effective diffusion-like 
equation for the large scales, which defines an eddy viscosity. 
A change of sign of the eddy viscosity corresponds to a change of the stability 
of the perturbation. In particular, the system becomes unstable when the eddy 
viscosity becomes negative \citep{sivashinsky1985negative,dubrulle1991eddy}.

The choice of the multiple-scale method to study the stability of the
dusty Kolmogorov flow is motivated by the fact that in the Newtonian
case (at $\phi=0$) the most unstable perturbation is indeed at large
scale, and the multiple-scale prediction for the critical Reynolds
number is correct.  For simplicity of the calculation, and in analogy
with the Newtonian case, we also assume that the most unstable
perturbation is transverse, i.e. depends on the large-scale variable
$X$ only and not on $Y$.  The validity of these assumptions for the
dusty gas at $\phi > 0$ will be checked by extensive numerical
simulations of the linear systems in Section \ref{sec:5}. In
particular, we anticipate that while the transverse nature of the
most unstable perturbation was always confirmed, in certain parameters
region the scale separation appears to be violated, when this happens
the multiscale approach is not providing the correct prediction.

Before proceeding, it is convenient to rewrite Eq.~(\ref{eq:psi-lin}) 
in terms of a co-stream function defined as $\chi=\Psi+\phi \Psi_p$. 
Such a choice allows to remove the apparent singularity of the 
Stokes drag in the limit $\tau \to 0$ in Eq.~(\ref{eq:psi-lin}). 
Indeed, by combining Eq.~(\ref{eq:psi-lin}) and Eq.~(\ref{eq:psip-lin}) we obtain
\begin{eqnarray}
\partial_t \nabla^2 \chi &+& U \cos(K y) (K^2+\nabla^2)\partial_x \chi-
\phi U K \left(K \cos(K y) \partial_y \Phi_p +\sin(K y) \nabla^2 \Phi_p \right)\nonumber\\
&-& \nu \nabla^4 (\chi-\phi \Psi_p) = 0
\label{eq:chi-lin}
\end{eqnarray}
which removes the explicit dependence of (\ref{eq:psi-lin}) on $\tau$.
The linear systems for the perturbative analysis is hence formed by the set of equations
(\ref{eq:chi-lin}), (\ref{eq:psip-lin}) and (\ref{eq:phip-lin}). 

Following the multiple scale method, we assume a perturbative
expansion of the fields:
\begin{eqnarray}
&& \chi(X,y,T) = \chi_0(X,y,T)+\varepsilon \chi_1(X,y,T)+\varepsilon^2 \chi_2(X,y,T) \nonumber \\
&& \Psi_p(X,y,T) = \Psi_{p,0}(X,y,T)+\varepsilon \Psi_{p,1}(X,y,T)+\varepsilon^2 \Psi_{p,2}(X,y,T) \label{eq:ms1} \\
&& \Phi_p(X,y,T) = \Phi_{p,0}(X,y,T)+\varepsilon \Phi_{p,1}(X,y,T)+\varepsilon^2 \Phi_{p,2}(X,y,T) \nonumber.
\end{eqnarray}
The derivative operators are transformed as 
$\partial_x \to \varepsilon \partial_X$, $\partial_t \to \varepsilon^2 \partial T$. 
Notice that the base flow does not depend on $x$ and $t$ 
and therefore the same holds for the perturbation.
By inserting the expansions (\ref{eq:ms1}) and the fast/slow 
variables decomposition into Eqs.~(\ref{eq:chi-lin}),(\ref{eq:psip-lin}) 
and (\ref{eq:phip-lin}) we obtain, at order $\varepsilon^0$, 
that the zero-order fields do not depend on the fast variable, 
i.e. $\chi_0(X,y,T)=a_0(X,T)$, $\Psi_{p,0}(X,y,T)=b_0(X,T)$ 
and $\Phi_{p,0}(X,y,T)=c_0(X,T)$. At the order $\varepsilon^2$, 
the absence of secular terms requires $c_0=0$.

The solvability condition is obtained by integrating 
Eqs.~(\ref{eq:psip-lin}-\ref{eq:chi-lin}) 
over one period of the fast variable $y$. 
The first non-trivial condition is obtained at order $\varepsilon^3$
and gives a relation among the large-scale fields 
\begin{equation}
a_0(X,T)=(1+\phi) b_0(X,T) \, .
\label{eq:ms1b}
\end{equation}
At order $\varepsilon^4$, 
we finally get the diffusion equation for the slow field $a_0$
\begin{equation}
\frac{2}{Re} (1+\phi) \partial_T \partial_X^2 a_0(X,T) + 
\nu Re \left(1 - \frac{2}{ Re^2} + 2 \phi + \phi^2 - 
\phi St \right) \partial_X^4 a_0(X,T) = 0\;,
\label{eq:ms2}
\end{equation}
which defines the eddy viscosity
\begin{equation}
\nu_e = \nu \frac{Re^2}{2(1+\phi)} \left(\frac{2}{Re^2} -(1+\phi)^2+ \phi St \right)\,.
\label{eq:ms3}
\end{equation}
The critical Reynolds number is finally obtained by the condition 
$\nu_e=0$ at which the eddy viscosity becomes negative, 
indicating that the basic flow is linearly unstable 
\citep{sivashinsky1985negative,dubrulle1991eddy}
\begin{equation}
Re_c = \sqrt{\frac{2}{(1+\phi)^2 - \phi St}}\,. 
\label{eq:ms4}
\end{equation}
For $\phi=0$ the critical Reynolds number predicted by Eq.~(\ref{eq:ms4}) 
recovers correctly the Newtonian value $Re_c=\sqrt{2}$. 
Interestingly, the same value is recovered also on the {\it neutral curve} 
$St=2+\phi$. For $St \to 0$ we obtain the Saffman limit 
$R_c=\sqrt{2}/(1+\phi)$. For $St<2$, Eq.~(\ref{eq:ms4}) predicts a 
monotonic decrease of $Re_c$ as a function of $\phi$ ($Re_c(\phi) \le Re_c(0)$) 
indicating that tiny particles always destabilize the flow. 
On the contrary, for $St>2$ the critical $Re$ depends on $\phi$ in a 
non-monotonic way: for $\phi < \phi_{max}=(St-2)/2$, $Re_c$ increases 
monotonically above the Newtonian value $\sqrt{2}$, it reaches a 
maximum at $\phi_{max}$ after which it monotonically decreases and for 
$\phi>St-2$ it goes below $\sqrt{2}$. Therefore, increasing the mass 
fraction particles first stabilize the flow up to a maximum then 
the stabilizing effect decreases and, finally, for large enough mass 
fraction particles make the flow more unstable than the in the Newtonian case. 
We remark that according to Eq.~(\ref{eq:ms4}) the dusty Kolmogorov flow 
should always be stable (i.e. $Re_c \to \infty$) for 
$St \ge (1+\phi)^2/\phi \ge 4$. 
We will see that this is actually an overestimation of the stability 
due to the fact that the main assumption of the multiple-scale analysis 
(instability to large-scale perturbations) does not 
hold in a certain region of the parameter space ($\phi$,$St$).

%%%%%%%%%%%%%%%%%%%%%%%%%%%%%%%%%%%%%%%%%%%%
\section{Numerical analysis}
\label{sec:5}

To check the validity of the analytical result (\ref{eq:ms4}) 
obtained with the multiple-scale analysis, 
we performed an extensive numerical study 
of the linearized equations in two dimensions 
(\ref{eq:psi-lin}-\ref{eq:phip-lin}) by means of a pseudo-spectral 
method in a square domain of size $L =2\pi $ with periodic boundary 
conditions. For each set of values of the parameters $\phi$ and $St$ 
in the range $0 \le \phi \le 6$ and $0 \le St \le 6$ we have studied 
the stability of the system at varying the $Re$ number. The latter 
has been varied by changing the amplitude of the forcing $F$ while 
keeping fixed the viscosity $\nu=10^{-3}$ and the scale of the base 
flow $1/K$. Simulations have been done at two different resolutions, 
with $128^2$ and $256^2$ grid points and forcing wavenumber $K=32$ 
and $K=64$, respectively, to check finite size effects. 
A random initial perturbation has been imposed to each Fourier mode 
$(k_x,k_y)$ in the range $ 0 \le |{\bm k}| \le K$.
The stability of each mode and its growth rate is determined 
by the temporal evolution of its amplitude after a short transient. 
The critical Reynolds number was determined by means of the 
bisection method based on the total kinetic energy of the fluid.

%%%%%%%%%%%%%%%%%%%%%%%%%%%%%%%%%%%%%%%%%%%%
\begin{figure}
\centerline{\includegraphics[width=1\textwidth]{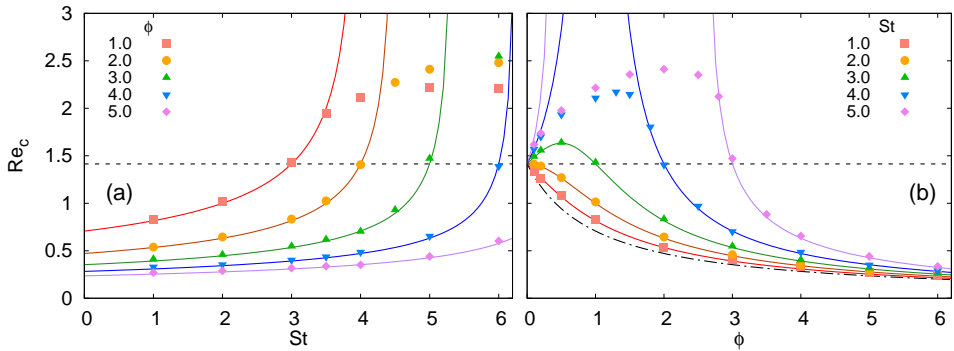}}
\caption{Critical Reynolds number (a) as a function of the Stokes 
number $St$ for different values of $\phi$ and (b) as a function of 
the mass fraction $\phi$ for different values of $St$. The values of 
the parameters $\phi$ and $St$ are reported in the legend. 
Solid curves denote the multiscale prediction (\ref{eq:ms4}); 
symbols the numerical results; dashed lines display the Newtonian 
value $\sqrt{2}$; dash-dotted line in panel (b) shows the Saffman 
limit $\sqrt{2}/(1+\phi)$.}
\label{fig:1}
\end{figure}
%%%%%%%%%%%%%%%%%%%%%%%%%%%%%%%%%%%%%%%%%%%%

In Figure~\ref{fig:1}(a) we plot the critical Reynolds number as a 
function of the Stokes number for different mass fraction values $\phi$. 
At small $St$, $Re_c$ is smaller than that of the single-phase fluid 
($Re_c=\sqrt{2}$ for $\phi=0$) and the numerical results are in agreement 
with the theoretical prediction (\ref{eq:ms4}). 
In particular, in the limit $St \ll 1$, the critical Reynolds number recovers the 
Saffman limit $Re_c = \sqrt{2}/(1+\phi)$ (not shown). 
The critical Reynolds number increases monotonically 
with $St$ at fixed $\phi$, eventually becoming larger than $\sqrt{2}$, 
meaning that large particle inertia has a stabilizing effect on the flow. 
At a qualitative level, 
the physical mechanisms of the stabilizing/destabilizing effect 
of the particles have been already discussed by \citet{saffman1962}. 
Particles with small $St$ follow the flow almost like tracers, 
so that their effect is simply to increase the density of the suspension. 
Therefore, the dusty gas behaves as a Newtonian flow with a reduced 
kinematic viscosity (see Eq.~\ref{eq:smalltau}) which makes the flow more unstable. 
Conversely, particles with large inertia do not follow the perturbation of the flow, 
but they "carry on with the velocity of the base flow"~\citep{saffman1962}. 
The disturbance has therefore to flow around the particles, 
dissipating its energy because of the viscous drag. 
Our numerical results show that the stabilizing effect at large $St$
is weaker than the prediction (\ref{eq:ms4}). 
In particular, the multiple-scale predicts unconditioned stability 
(i.e. $Re_c =\infty$) for $St \ge (1+\phi)^2/\phi$, 
while in the numerical simulations $Re_c$ remains finite. 

The behavior of $Re_c$ as a function of the mass fraction $\phi$ for 
fixed values of $St$, shown in figure~\ref{fig:1}(b), gives further 
insights on the stability of the system. In agreement with 
Eq.~(\ref{eq:ms4}), we find that $Re_c$ is monotonically decreasing 
for $St \le 2$ (i.e. tiny particles always destabilize the flow). 
Conversely, at $St>2$ the particles at low concentration stabilize the 
flow while at sufficiently large concentrations $\phi \ge St - 2$ they 
have a destabilizing effect. 
It is interesting to note that a similar non-monotonic behavior as a function of the mass loading 
has been observed also for the skin-friction coefficient 
in Lagrangian-Eulerian simulations of inertial particles in a vertical channel flow 
\citep{capecelatro2018transition}. 
The physical mechanism of the destabilizing effect at large $\phi$ 
is similar to that of the case of small $St$. 
The strong drag exerted by the large mass fraction forces the fluid to follow closely the particle velocity
(see Eq.~\ref{eq:largephi1}).  
As a consequence the dusty gas behaves almost as a single-phase fluid with a larger density 
and therefore a smaller kinematic viscosity, which reduces its stability. 
From Fig.~\ref{fig:1}(b) it is evident 
that the agreement between the multiple-scale result and numerical 
simulations is very good for any $\phi$ up to $St=3$. For $St \ge 4$, 
the multiple-scale result (\ref{eq:ms4}) overestimates the $Re_c$ for 
an intermediate interval of values of $\phi$ around $\phi_{max} = (St-2)/2$. 
Nonetheless, also in these cases ($St=4$ and $St=5$) 
the multiple-scale prediction works well for small and large values of $\phi$.

%%%%%%%%%%%%%%%%%%%%%%%%%%%%%%%%%%%%%%%%%%%%
\begin{figure}
\centerline{\includegraphics[width=1\textwidth]{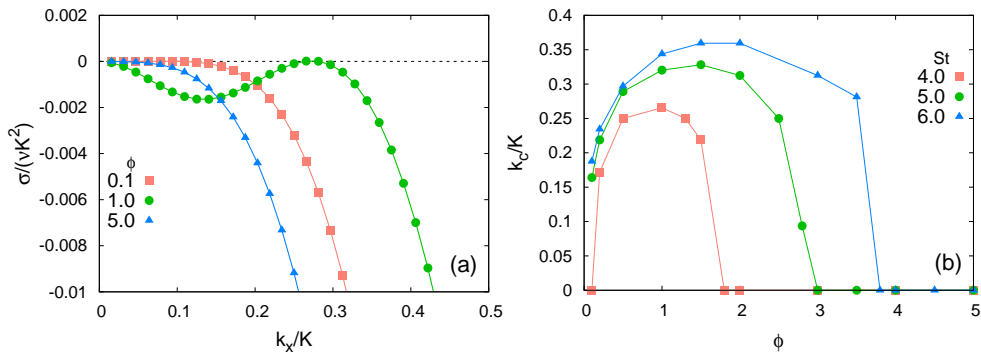}}
\caption{Panel (a): Non-dimensional growth rate $\sigma/(\nu K^2)$ 
as a function of the $x$-wavenumber $k_x$ at the onset of the instability 
$Re=Re_c$ for $St=4$ and different values of the mass fraction $\phi$ 
as labeled. Panel (b): First unstable mode $k_c$ normalized by $K$ as 
a function of $\phi$ for different values of $St$ as labeled.}
\label{fig:2}
\end{figure}
%%%%%%%%%%%%%%%%%%%%%%%%%%%%%%%%%%%%%%%%%%%%

In order to understand why the multiple-scale analysis fails in 
predicting the correct $Re_c$ at large $St$, we computed numerically 
the growth rate $\sigma$ as a function of the wavenumber $k_x$ of the 
perturbation (i.e. the dispersion relation) for different 
values of the parameters $\phi$ and $St$. In Figure~\ref{fig:2}(a) we 
show the dispersion relation computed at the critical point $Re=Re_c$ 
for $St=4$ and three values of the mass fraction. For small and large 
mass fraction ($\phi =0.1$ and $\phi =5$) we observe that the growth 
rate $\sigma$ is a monotonically decreasing function of $k_x$ and the 
unstable mode is the smallest available wavenumber 
$k_c = k_{min} \equiv 2\pi/L$. In these cases, the hypothesis of large scale 
separation is justified and indeed the predictions of the 
multiple-scale analysis are in agreement with the numerical results. 
Conversely, for $\phi=1$ the curve $\sigma(k_x)$ is non-monotonic and 
the unstable mode appears to be at $k_c \simeq 0.3 K$, therefore
the instability is no more triggered by large-scale perturbations
and multiple scale analysis fails to predict the instability.
Similar behaviors have been observed also for $St=5$ and 
$St=6$ (not shown). To systematically investigate the region of 
parameters for which the multiple-scale analysis is not expected to 
work we numerically studied the dependence of the unstable 
(transverse) mode $k_c$ on $\phi$ and $St$ at $Re=Re_c$, 
shown in Figure~\ref{fig:2}(b). 
For $St \ge 4$ and intermediate values of $\phi$ we found $k_c \simeq 0.3 K$, 
while for $St \le 3$ (not shown) we always found $k_c=k_{min}$ 
in agreement with the multiple-scale assumption. 
By comparing Figures~\ref{fig:2}(b) and ~\ref{fig:1}(b) we clearly observe 
the correspondence between the theoretical-numerical agreement 
in Figure~\ref{fig:1}(b) and the fact that $k_c \ll K$.

%%%%%%%%%%%%%%%%%%%%%%%%%%%%%%%%%%%%%%%%%%%
\section{Conclusions}
\label{sec:conclusions}

\begin{figure}
\center
\includegraphics[width=0.7\textwidth]{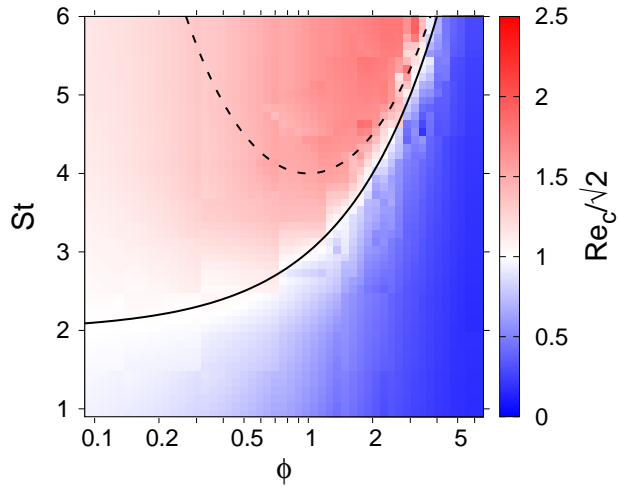}
\caption{Critical Reynolds number as a function of $\phi$ and $St$. 
Red (blue) color-scale denotes regions more stable (unstable) than 
the Newtonian flow in which $Re_c > \sqrt{2}$ ($Re_c < \sqrt{2}$). 
Solid black line is the neutral curve $St=\phi+2$ at which $Re_c=\sqrt{2}$. 
Dashed line represents the border of the region of unconditioned stability 
predicted by the multiple-scale analysis $St > (1+\phi)^2/\phi$.}
\label{fig:3}
\end{figure}

We have investigated the linear stability of a dilute suspension of heavy particles 
in the Kolmogorov flow within the Eulerian model proposed by \citet{saffman1962}.
In the absence of particles, it is well known that the value of the critical Reynolds 
number $Re_c =\sqrt{2}$ for the stability of the laminar base flow can be obtained 
by means of a multiple-scale analysis. Here we have adopted the same approach to 
extend the study of the linear stability to the full parameter space of the Saffman 
model given by the Reynolds number $Re$, the mass fraction $\phi$ and the 
Stokes number $St$. The multiple-scale prediction for the onset of the instability, 
$Re_c = \sqrt{2/((1+\phi)^2-\phi St)}$, as a function of $St$ and $\phi$ has been 
compared with the results of numerical simulations of the linearized system. 
Figure~\ref{fig:3} summarizes the main results.
%%%%%%%%%%%%%%%%%%%%%%%%%%%%%%%%%%%%%%%%%%%%
%%%%%%%%%%%%%%%%%%%%%%%%%%%%%%%%%%%%%%%%%%%%
Particles with small inertia ($St < \phi+2$, blue region) reduce the stability 
of the base laminar flow. Conversely, the presence of particles with large 
inertia ($St > \phi+2$, red region) retard the onset of the instability. 
The prediction of the neutral curve $St = \phi+2$ in which the effect 
of the particles on the linear stability vanishes is confirmed by numerics. 
In general, we have found that the multiple-scale analysis correctly predicts 
the values of $Re_c$ in a large part of the parameter space. 
It correctly recovers the limit of a Newtonian flow with rescaled viscosity 
$\nu/(1+\phi)$ both for $St \ll 1$ and $\phi \gg 1$. 
Nonetheless, for large $St$ it overestimates $Re_c$ in an intermediate 
range of $\phi$. In particular, the region of unconditioned stability 
$St > (1+\phi)^2/\phi$ is not observed in the numerics. 
By investigating numerically the dispersion relation at the critical Reynolds number, 
we have found that the failure of the multiple-scale prediction is due to 
the lack of scale separation between the most unstable mode and the 
wavenumber of the base flow, thus invalidating the assumptions of the 
perturbative approach in that parameter region.
A natural extension of the present work would be to investigate 
the weakly non-linear dynamics of the Kolmogorov-Saffman system 
and the structure of the secondary flow above $Re_c$ \citep{sivashinsky1985weak}.

We conclude with two comments concerning the choice of the Kolmogorov base flow 
and the Saffman model. The first is related to the preferential concentration of inertial particles, 
which, in principle, can be observed also in laminar flow. 
For a parallel flow (such as Kolmogorov one), the fixed point solution of the model 
has a uniform particle density field and the infinitesimal perturbation of the density 
is passively transported in the linearized dynamics. Therefore, preferential concentration 
does not influence the linear stability of the Saffman model. 
To investigate such effects requires the choice of a different base flow.

The second concerns the relevance of our results to real-world systems. 
Modeling the coupling between the particles and the fluid in particle-laden flows 
is a challenging task which requires a compromise between accuracy and simplicity. 
The simplicity of the Saffman model combined with that of the Kolmogorov flow allowed us 
to obtain an analytic prediction for the critical Reynolds number. 
Our results could, in principle, differ quantitatively from those of more refined models 
(e.g., Lagrangian models with accurate implementation of the two-way coupling). 
Nonetheless, our work offers a qualitative benchmark for future experimental studies 
and numerical simulations based on Lagrangian approaches which allow to include additional effects, 
such as finite particle size or particle-particle interactions, which are not captured by the Saffman model. 
The comparison between our results and those obtained by means of more accurate models 
could improve our understanding concerning the impact of such complex processes on the stability of laminar flows.

\vspace{-0.2cm}

%%%%%%%%%%%%%%%%%%%%%%%%%%%%%%%%%%%%%%%%%%%%
\section{Acknowledgments}
We acknowledge HPC CINECA for computing resources 
(INFN-CINECA Grant No. INFN20-FieldTurb). 
G.B. and S.M. acknowledges support from the Departments of Excellence 
grant (MIUR). A.S. acknowledges the Italian research project MODSS 
(Monitoring Debris Space Stereo) Grant No. ID 85-2017-14966, funded 
by Lazio Innova (Regione Lazio).

\vspace{-0.2cm}

\section{Declaration of Interest}
The authors report no conflict of interest.

\vspace{-0.2cm}

%%%%%%%%%%%%%%%%%%%%%%%%%%%%%%%%%%%%%%%%%%%%
\appendix

%%%%%%%%%%%%%%%%%%%%%%%%%%%%%%%%%%%%%%%%%%%%
\section{Squire's theorem for the Saffman model}
\label{app:a}
We consider a generic parallel basic flow ${\bm U}=(U(z),0,0)$ in a 
three-dimensional domain. The linearized Saffman model around the 
basic flow (\ref{eq:u-lin}-\ref{eq:v-lin}) written in non-dimensional form is 
\begin{equation}\label{eq:a1}
\partial_t {\bm u} + ({\bm U}\cdot{\bm \nabla}) {\bm u} + 
({\bm u}\cdot{\bm \nabla}) {\bm U} = 
- {\bm \nabla}p + {1 \over Re} \nabla {\bm u} + {\phi \over Re St} 
\left( {\bm v} - \bf{u} \right)
\end{equation}
\begin{equation}\label{eq:a2}
\partial_t {\bm v} + ({\bm U}\cdot{\bm \nabla}) {\bm v} + 
({\bm v}\cdot{\bm \nabla}) {\bm U} = 
- {1 \over Re St} \left( {\bm v} - \bf{u} \right)
\end{equation}
where $Re=U /(\nu K)$ and $St=\tau \nu K^2$ and 
$1/K$ is the characteristic scale of $U(z)$. 
We now perform a Fourier transform in the directions $x$, $y$ and $t$ and write
$\left\{ {\bm u}, {\bm v}, p \right\} = 
\left\{ {\widehat {\bm u}}(z), {\widehat {\bm v}}(z), {\widehat p}(z) \right\} 
\exp(i {\bm k}_h\cdot{\bm x}_h - i \omega t )$,
where $\bm x_h=(x,y)^T$ and $\bm k_h=(k_x,k_y)^T$, with $T$ denoting the transpose.
Introducing the notation
${\bm U}_h = \left(U(z), 0\right)^T$, 
${\widehat {\bm u}}_h = \left(\widehat{u}_x,\widehat{u}_y\right)^T$, 
and ${\widehat {\bm v}}_h = \left(\widehat{v}_x, \widehat{v}_y\right)^T$, 
the linearized equations in normal modes take the form
\begin{equation}\label{eq:a6}
(-i \omega + i {\bm k}_h\cdot{\bm U}_h)\widehat{\bf{u}}_h + 
\widehat{\bf{u}}_z {d {\bm U}_h \over dz} 
= - i {\bm k}_h \widehat{p} + 
{1 \over Re} \left({d^2 \over z^2}-{\bm k}_h^2 \right) \widehat{\bf{u}}_h 
+ {\phi \over Re St} \left( \widehat{\bf{v}}_h - \widehat{\bf{u}}_h \right)
\end{equation}
\begin{equation}\label{eq:a7}
(-i \omega + i {\bm k}_h\cdot{\bm U}_h)\widehat{u}_z 
= - {d\widehat{p} \over dz} + 
{1 \over Re} \left({d^2 \over dz^2}-{\bm k}_h^2 \right) \widehat{u}_z 
+ {\phi \over Re St} \left( \widehat{v}_z - \widehat{u}_z \right)
\end{equation}
\begin{equation}\label{eq:a8}
(-i \omega + i {\bm k}_h\cdot{\bm U}_h)\widehat{\bf{v}}_h + 
\widehat{\bf{v}}_z {d {\bm U}_h \over dz} 
= -{1 \over Re St} \left( \widehat{\bf{v}}_h - \widehat{\bf{u}}_h \right)
\end{equation}
\begin{equation}\label{eq:a9}
(-i \omega + i {\bm k}_h\cdot{\bm U}_h)\widehat{v}_z 
= - {1 \over Re St} \left( \widehat{v}_z - \widehat{u}_z \right)
\end{equation}
The linearized dynamics described by the Eqs.~(\ref{eq:a6}-\ref{eq:a9}) 
is independent for each mode ${\bf k}_h$. 
Therefore, for each mode ${\bf k}_h$ it is possible to perform a rotation 
of the Fourier amplitudes of the velocity fields $\widehat{\bm u}_h$ and 
$\widehat{\bm v}_h$ in the direction of the wave-vector ${\bf k}_h$ 
by means of the following transformation:
\begin{equation}\label{eq:a10}
\begin{array}{c}
\overline{k}_x = |{\bm k}_h|, \quad 
\overline{\omega} = { \overline{k}_x \over k_x} \omega, \quad
\overline{Re} = {k_x \over \overline{k}_x} Re \le Re, \quad \\[0.5cm]
\overline{u}_x = { {\bm k}_h \cdot {\widehat {\bm u}}_h \over |{\bm k}_h|}, \quad 
\overline{u}_z = \widehat{u}_z, \quad 
\overline{p} = {\overline{k}_x \over k_x} \widehat{p}, \quad
\overline{v}_x = { {\bm k}_h \cdot {\widehat {\bm v}}_h \over |{\bm k}_h|}, \quad 
\overline{v}_z = \widehat{v}_z, \quad 
\end{array}
\end{equation}
From Eqs.(\ref{eq:a6}-\ref{eq:a9}) one obtains the equations for the new variables
\begin{equation}\label{eq:a11}
\left[ - i \overline{\omega} + i \overline{k}_x U \right] \overline{u}_x + 
\overline{u}_z {d U \over dz} = 
- i \overline{k}_x \overline{p} + 
{1 \over \overline{Re}} \left( {d^2 \over dz^2} - \overline{k}_x^2 \right) \overline{u}_x 
+ {\phi \over \overline{Re} St} \left( \overline{v}_x - \overline{u}_x \right)
\end{equation}
\begin{equation}\label{eq:a12}
\left[ - i \overline{\omega} + i \overline{k}_x U \right] \overline{u}_z = 
- {d\overline{p} \over dz} + 
{1 \over \overline{Re}} \left( {d^2 \over dz^2} - \overline{k}_x^2 \right) \overline{u}_z 
+ {\phi \over \overline{Re} St} \left( \overline{v}_z - \overline{u}_z \right)
\end{equation}
\begin{equation}\label{eq:a13}
\left[ - i \overline{\omega} + i \overline{k}_x U \right] \overline{v}_x + \overline{v}_z {d U \over dz} = 
- {1 \over \overline{Re} St} \left( \overline{v}_x - \overline{u}_x \right)
\end{equation}
\begin{equation}\label{eq:a14}
\left[ - i \overline{\omega} + i \overline{k}_x U \right] \overline{v}_z = 
- {1 \over \overline{Re} St} \left( \overline{v}_z - \overline{u}_z \right)
\end{equation}
The new system of equations (\ref{eq:a11}-\ref{eq:a13}) 
is formally identical to the original one (\ref{eq:a6}-\ref{eq:a9}) 
in which one imposes a purely two-dimensional perturbation 
with $\widehat{u}_y = \widehat{v}_y= 0$ and $k_y=0$. 
Therefore, three-dimensional perturbations which are unstable 
at a given $Re$ correspond to two-dimensional disturbance at 
smaller Reynolds number $\overline{Re}$ 
(and at the same $\phi$ and $St$) with larger growth rate 
($\Im(\overline{\omega}) \ge \Im(\omega) > 0$).

%%%%%%%%%%%%%%%%%%%%%%%%%%%%%%%%%%%%%%%%%%%%%

\bibliographystyle{jfm}
\bibliography{biblio}

\end{document}